\documentclass[twocolumn,showpacs,preprintnumbers,amsmath,amssymb]{revtex4}
\usepackage{cases}
\usepackage{amssymb}
\usepackage{txfonts}
\usepackage{amsmath}

\usepackage{graphicx}
\usepackage{dcolumn}
\usepackage{bm}

\begin{document}
\title{Broadband enhancement of light harvesting in luminescent solar concentrator}
\author{Yun-Feng Xiao$^{1}$\footnote{To whom correspondence should be addressed.
\newline Email: yfxiao@pku.edu.cn. Phone: +86(10)62754867.\newline URL: http://www.phy.pku.edu.cn/~yfxiao/index.html},
Chang-Ling Zou$^{2}$, Yi-Wen Hu$^{1}$, Yan Li$^{1}$, Lixin Xiao$^{1}$, Fang-Wen Sun$^{2}$, and Qihuang Gong$^{1}$}%

\affiliation{$^{1}$State Key Lab for Mesoscopic Physics, School of Physics, Peking
University, Beijing 100871, P. R. China}
\affiliation{$^{2}$Key Lab of Quantum Information, University of Science and Technology of
China, Hefei 230026, Anhui, P. R. China}

\begin{abstract}
\noindent Luminescent solar concentrators (LSCs) are large-area devices that
may absorb incident sunlight, then emit luminescence photons with high quantum
efficiency which finally be collected by a small photovoltaic (PV) system. The
light-harvesting area of the PV system is much smaller than that of the LSC
system, potentially reducing the cost of solar cells. Here, we present a
theoretical description of the luminescent process in nanoscale LSCs where the
conventional ray-optics model is no longer applicable. We demonstrate that a
slot waveguide consisting of a nanometer-sized low-index slot region
sandwiched by two high-index regions provides a broadband enhancement of light
harvesting by the luminescent centers in the slot region. This is because the
slot waveguide can (1) greatly enhance the spontaneous emission due to the
Purcell effect, (2) dramatically increase the effective absorption length of
luminescent centers, and (3) strongly improve the fluorescence quantum yield
of luminescent centers. It is found that about $80\%$ solar photons can be
re-emitted even for a low fluorescent quantum yield of $0.5$, and $80\%$
re-emitted photons can be coupled to the slot-waveguide. This LSC is potential
to construct a tandem structure which can absorb nearly full-spectrum solar
photons, and also may be of special interest for building integrated
nano-solar-cell applications.

\end{abstract}
\date{\today }
\maketitle

In the past few years, many approaches involving nanostructures or
nanostructured materials have been proposed to reduce cost and improve
efficiency in both experiment \cite{nano3,nano4,nano8,nano12} and theory
\cite{t2,t5,t6}. On the other hand, concentrators with large-area optical
components to collect direct sunlight and transfer the energy to small,
high-efficiency photovoltaic (PV) cells have been suggested as a simple
approach to lower the cost per peak Watt of solar cell systems for many
decades \cite{jason}. To overcome the excess heat problem, chromatic
aberrations and expensive maintaining in these imagining concentrators,
luminescent solar concentrators (LSCs) represent an alternative approach to
lower the costs of solar cell systems
\cite{batchelder,weber,AP1977,michael,IEEE}. LSCs generally consist of
low-cost transparent sheets doped with luminescent species, such as dye
molecules and quantum dots. Incident sunlight is highly absorbed by the
luminescent centers and luminescence is emitted with high fluorescence quantum
yield (FQY, defined as the ratio of the number of photons emitted to the
number of photons absorbed), so that the emitted photon is trapped in the
sheet by total internal reflection and travels to the edges where it can be
collected by solar cells. The active material layer can be much thinner than
the intrinsic absorption length of the material, thus dramatically reducing
the amount of the solar cell material.

One of the key parameters of LSCs is the coupling efficiency $\beta$ which
describes the ratio of luminescent photons coupled to waveguide modes for
ultimate collection by PV systems on the waveguide edge. To model these LSCs
devices, thermodynamical and computational ray tracing approaches have been
introduced, and both approaches represent a broad-scale, macroscopic model
\cite{michael,Chatten,Earp,Carrascosa,Rau}. For example, conventional LSCs
have a thickness much larger than the wavelength, so that the ray-optics model
can be applied to estimate the collection efficiency of luminescence by the
solar cell. However, with LSCs devices moving to the nanoscale, the ray-optics
picture and some basic assumptions are no longer strictly applicable. In this
paper, we propose a nanometer-sized slot waveguide as the main structure of
LSCs, and theoretically study the enhancement of spontaneous emission
(described by the factor $F_{p}$) of luminescent centers in this slot
waveguide based on the Fermi-golden rule. Remarkably, this great enhancement
of spontaneous emission predicts not only an increased absorption length of
luminescent centers but also a very large waveguide coupling efficiency
$\beta$. We demonstrate that such a slot waveguide LSC provides a broadband
enhancement of light harvesting. It is found that about $80\%$ solar photons
can be re-emitted even for a low initial FQY of $0.5$, and $80\%$ re-emitted
luminescent photons can be coupled into the slot waveguide modes for ultimate
collecting by the solar cell located on the waveguide edge. This LSCs may be
of special interest for building integrated solar cell applications.

\begin{figure}[ptbh]
\centerline{\includegraphics[keepaspectratio=true,width=0.45\textwidth]{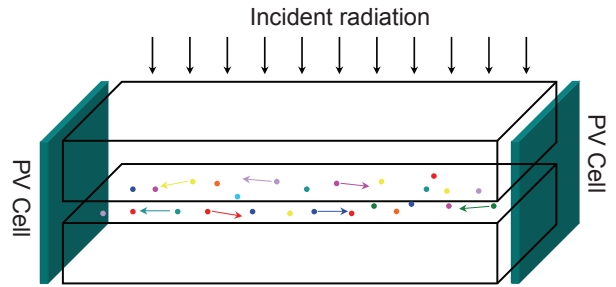}}\caption{Schematic
illustration of a slot waveguide structured LSC. The slot waveguide consists
of a nanometer-sized low-index slot region sandwiched by two high-index
regions. Active medium such as dye molecules and quantum dots are located in
the slot region. The PV system is on the edge of the slot waveguide and can
collect the luminescent photons coupled in the slot waveguiding mode.}%
\end{figure}

The basic structure of the proposed design is shown in Figure 1 and the
cross-sectional geometry is depicted in Figure 2(b). The slot waveguide
consists of a nanometer-sized low-index slot region (with the permittivity
$\varepsilon_{1}=n_{1}^{2}$, here $n_{1}$ is assumed $1.4$) sandwiched by two
high-index transparent layers (with the permittivity $\varepsilon_{2}%
=n_{2}^{2}$, here $n_{2}$ is assumed $2.4$), and the PV systems are on the
edge of the slot waveguide. In this work, high-index polymer or glass layers
are desirable to indeed reduce the cost, but it requires future studies.
Active medium such as dye molecules and quantum dots are doped in the slot
region. The active medium absorbs the incident sunlight and re-emits the
luminescence with a red-shifted wavelength. Different from the conventional
waveguides, slot waveguide is able to guide and strongly confine light in a
nanoscale low-index material. Quantitatively, the mode area of the slot
waveguide can be reduced to at least $1/3$ of the conventional one. For
instance, the effective mode areas are about $0.008$ $\mathrm{\mu m}^{2}$ and
$0.022$ $\mathrm{\mu m}^{2}$ for waveguides in Figure 2 with and without the
slot, respectively. The basic principle of slot waveguide is based on the
discontinuity of the electric field at a normal boundary between two materials
\cite{OL2004}. When the electromagnetic wave propagates along the waveguide
direction, the major component of the electric field of the quasi-transverse
electric (TE) mode undergoes a discontinuity at the slot interfaces. According
to the boundary condition of Maxwell's equations, the amplitude of the
electric field in the low-index slot is much larger than that in the
high-index waveguides, and the ratio between them is $\varepsilon
_{2}/\varepsilon_{1}$. With a full-vectorial finite element method (FEM), we
can simulate the modes in the structure. As shown in Figure 2, the electric
field of the quasi-TE mode is strongly concentrated in the slot region and it
is normal to the interface, while the electric field of the
quasi-transverse-magnetic (TM) mode distributes over the whole cross-sectional
area and it is parallel to the interface. This unique characteristic makes the
slot waveguide attractive for numerous applications, for instance, highly
sensitive biosensors \cite{OL2007} and waveguide-based light source
\cite{OE2009}. In this paper, also due to the large electric field intensity
of the slot waveguide mode, the luminescent photons are expected to couple to
the propagating slot mode, and finally be collected by the smaller
high-efficiency PV systems. Due to the concentration effect at the LSCs edges,
the amount of the solar cell material reduces dramatically, and the cost of
solar cell may decrease greatly with the help of the slot waveguide LSC.
\begin{figure}[ptbh]
\centerline{\includegraphics[keepaspectratio=true,width=0.5\textwidth]{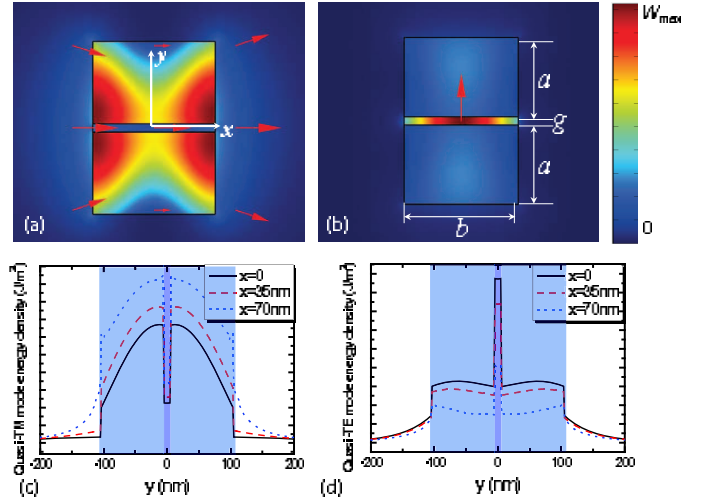}}\caption{(a),
(b) False-color representations of electromagnetic energy density
distributions $W(\vec{r})=\varepsilon_{0}d(\varepsilon(\vec{r})\omega
)/d\omega|\vec{E}(\vec{r})|^{2}+\mu_{0}|\vec{H}(\vec{r})|^{2}$ for quasi-TM
and -TE modes, respectively, where the red arrows show the directions of the
electric fields. The geometry of the slot waveguide is depicted in (b). (c),
(d) Distributions of the normalized energy density $W(\vec{r})$ along the $y$
direction at several typical $x$ positions for quasi-TM and -TE modes,
respectively. Here, $\lambda=700$ nm; $a=100$ nm, $b=150$ nm and $g=10$ nm;
$n_{1}=1.4$, $n_{2}=2.4$.}%
\end{figure}
To analyze this LSC system, we note that the conventional ray-optics model is
not applicable because the thickness of the slot region is nanometer-sized.
Thus, we derive a simple analytical formula from the Fermi-golden rule to
obtain the spontaneous emission enhancement $F_{p}$ and the waveguide coupling
ratio $\beta$. In the weak-coupling regime, the spontaneous emission rate of a
dipole can be calculated from

\begin{equation}
\gamma_{\mathrm{WG}}=2\pi\left\vert g(\vec{r})\right\vert ^{2}\rho(\omega),
\label{eq1}%
\end{equation}
where $\left\vert g(\vec{r})\right\vert $ denotes the coupling strength
between the dipole $\vec{d}$ and the electromagnetic field $\vec{E}$ at the
dipole position $\vec{r}$. $\rho(\omega)$ represents the density of states. If
we assume that the dipole is oriented parallel to the electric field, the
coupling strength is then given by
\begin{equation}
\left\vert g(\vec{r})\right\vert =\left\vert \vec{d}\cdot\vec{E}(\vec
{r})/\hbar\right\vert , \label{eq2}%
\end{equation}
with%
\begin{equation}
\vec{E}(\vec{r})=\sqrt{\frac{\hbar\omega}{2\varepsilon_{0}\varepsilon
_{1}A_{\mathrm{eff}}l}W(\vec{r})}\frac{\vec{E}(\vec{r})}{\left\vert \vec
{E}(\vec{r})\right\vert }. \label{eq3}%
\end{equation}
Here $W(\vec{r})$ designates the normalized electromagnetic energy density
distribution; $A_{\mathrm{eff}}$ defines the effective mode area as
\[
A_{\mathrm{eff}}=\iiint_{V}\varepsilon_{0}\varepsilon(\vec{r})\left\vert
\vec{E}(\vec{r})\right\vert ^{2}dV/\max\left[  \varepsilon_{0}\varepsilon
(\vec{r})\left\vert \vec{E}(\vec{r})\right\vert ^{2}\right]  ,
\]
which plays the role analogous to the mode volume in cavity QED; $l$ is an
arbitrary quantization length which can be canceled later. For a slot
waveguide with a sufficiently small cross-sectional area, it only supports a
single quasi-TE and a single quasi-TM mode. As mentioned before, the TE mode
has the maximum electric field in the slot region. Thus, assuming a
one-dimensional density of states, the density of states can be expressed as%
\begin{equation}
\rho(\omega)=\frac{l}{\pi v_{g}(\omega)}, \label{eq4}%
\end{equation}
where $v_{g}=c/n_{g}$ is the group velocity of the slot waveguide mode with a
group index $n_{g}$. Here, $c$ is the light velocity in vacuum. Combining
Equations (1)-(4), we can obtain the spontaneous emission rate of the dipole
in the slot waveguide%
\begin{equation}
\gamma_{\mathrm{WG}}=2\pi\left\vert \vec{d}\right\vert ^{2}W(\vec{r}%
)\frac{\omega}{2\hbar\varepsilon_{0}\varepsilon_{1}A_{\mathrm{eff}}}\frac
{1}{\pi v_{g}(\omega)}. \label{eq5}%
\end{equation}
With the spontaneous emission rate into free space $\gamma_{0}=d^{2}\omega
^{3}/(3\pi\hbar\varepsilon_{0}c^{3})$, the emission enhancement also known as
Purcell effect, is thus expressed as%
\begin{equation}
F_{p}\equiv\frac{\gamma_{\mathrm{WG}}}{\gamma_{0}}=\frac{3}{4\pi}\frac
{c}{v_{g}(\omega)}\frac{\left(  \lambda_{0}/n_{1}\right)  ^{2}}%
{A_{\mathrm{eff}}}W(\vec{r}). \label{eq6}%
\end{equation}
Clearly, if the effective mode area $A_{\mathrm{eff}}$ of the slot waveguide
mode is squeezed significantly below $\lambda_{0}^{2}$, the light-matter
interaction can be dramatically enhanced, thus leading to an enhanced $F_{p}$.
In addition, the group velocity in the slot waveguiding mode is reduced, which
can furthermore increase the local density of states and also contribute to
the increase of $F_{p}$. Once the emission enhancement is given, the waveguide
coupling efficiency $\beta$ describing the ratio of the emission coupled to
the slot waveguiding mode, can be calculated as%
\begin{equation}
\beta=\frac{F_{p}}{F_{p}+n_{1}}. \label{eq7}%
\end{equation}

\begin{figure}[ptbh]
\centerline{\includegraphics[keepaspectratio=true,width=0.45\textwidth]{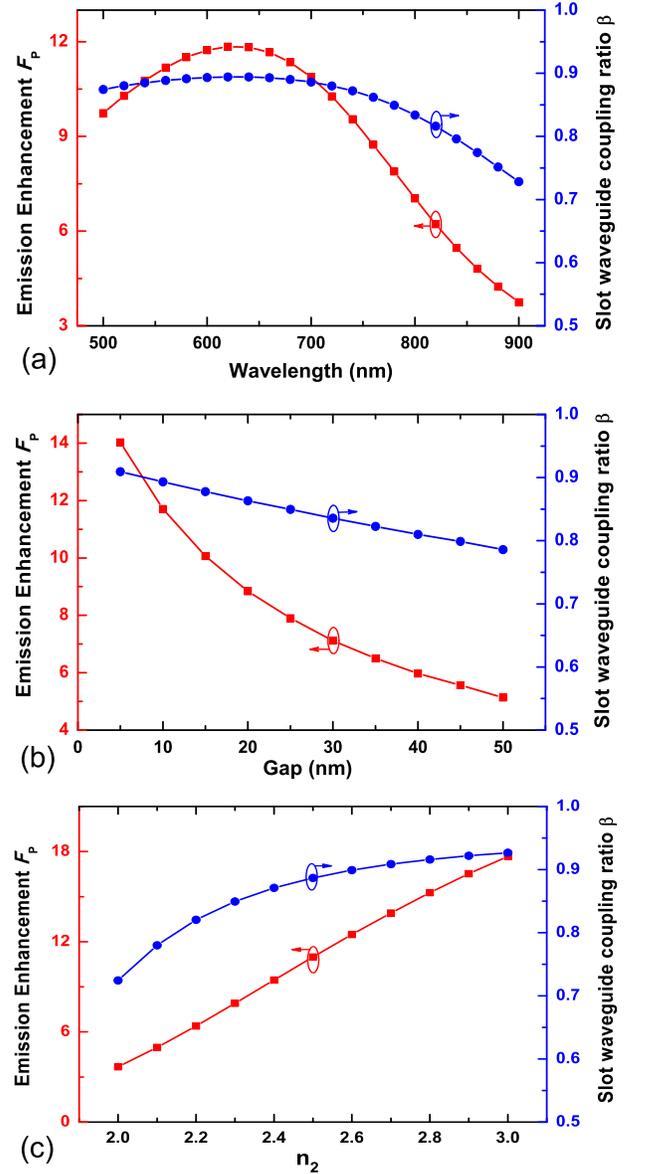}}\caption{(a)
Spontaneous emission enhancement ($F_{p}$) and waveguide coupling efficiency
($\beta$) depending on the working wavelength for slot waveguides with $g=10$
nm. (b) $F_{p}$ and $\beta$ depending on the slot width (gap) $g$ with
$\lambda=650$ nm. In all cases, $a=100$ nm, $b=150$ nm. (c) $F_{p}$ and
$\beta$ depending on $n_{2}$ with $\lambda=550$ nm, $g=10$ nm. In all cases,
$a=100$ nm, $b=150$ nm.}%
\end{figure}

Equations (6) and (7) represent the main physical parameters of the proposed
LSC. To numerically evaluate the emission enhancement and the waveguide
coupling efficiency, we still resort to the FEM simulation, as it can provide
not only the mode distribution ($f(\vec{r})$, $A_{\mathrm{eff}}$) but also the
group velocity $v_{g}(\omega)$ for a given geometry and working wavelength.
Figure 3(a) shows the calculated $F_{p}$ and $\beta$ as a function of
free-space wavelength $\lambda_{0}$ when a dipole is placed in the center of
the cross-section of the slot waveguide. We find two points. (i) $F_{p}$ first
increases and then decreases with the emission wavelength from $500$ to $900$
nm. This phenomenon is because that the effective mode area $A_{\mathrm{eff}%
}/\lambda_{0}^{2}$ first decreases and then increases. For the short
wavelength, most energy distributes in the high-index layers, and the field
enhancement effect in the slot becomes weaker. As a result, the effective mode
area increases. For the long wavelength, the small size of the slot waveguide
cannot support the guiding mode well. Therefore, more and more energy diffuses
outside for longer wavelength, and the effective mode area strongly expands.
(ii) $F_{p}$ exceeds $7$ over a broad wavelength range ($500-800$ nm). For
longer wavelengths, a larger-sized slot waveguide can be employed to reach the
maximum $F_{p}$, which will be discussed in the following. This strong
enhancement of the spontaneous emission takes advantage of nonresonance (large
bandwidth), which benefits from the subwavelength mode area (as small as
$0.02\left(  \lambda_{0}/n_{1}\right)  ^{2}$) and the reduced group velocity
(with the group index $n_{g}\sim2-2.7$) provided by the slot waveguide.
Remarkably, this broadband enhancement character overcomes the narrow
bandwidth limit of a microresonator-based emission enhancement. As a result,
most of the spontaneous emission can efficiently couple to the slot waveguide
mode, and $\beta$ exceed $0.8$ over a broad wavelength range ($500-800$
\textrm{nm}).

The enhancement of spontaneous emission also depends on the geometry of the
slot waveguide, for example, the slot width. Figure 3(b) shows the calculated
enhancement factor $F_{p}$ as a function of the slot width $g$. The factor
$F_{p}$ increases monotonically with the slot width reducing. For example,
$F_{p}$ increases from $5$ to $14$ when $g$ decreases from $50$ to $5$ nm. The
underlying physics is that both the effective mode area and the group velocity
are strongly reduced by decreasing the slot width. For instance,
$A_{\mathrm{eff}}$ and $n_{g}$ are $0.0223$ $\mathrm{\mu m}^{2}$ and $2.15$ at
$g=50$ nm, $0.0104$ $\mathrm{\mu m}^{2}$ and $2.82$ at $g=5$ nm. Thus, in this
slot width range, the waveguide coupling efficiency $\beta$ changes from
$0.78$ to $0.9$.

As mentioned above, we use $2.4$ as the default $n_{2}$. However, for a
different $n_{2}$, $F_{p}$ and $\beta$ can experience a significant change.
From Figure 3(c) we can see that with a larger $n_{2}$, both $F_{p}$ and
$\beta$ improve, which is because high-index-contrast interfaces can bring out
a better enhancement and confinement of light \cite{OL2004}. Therefore,
materials with both high transparency and refractive index are strongly
preferred in our work.

\begin{figure}[ptbh]
\centerline{\includegraphics[keepaspectratio=true,width=0.45\textwidth]{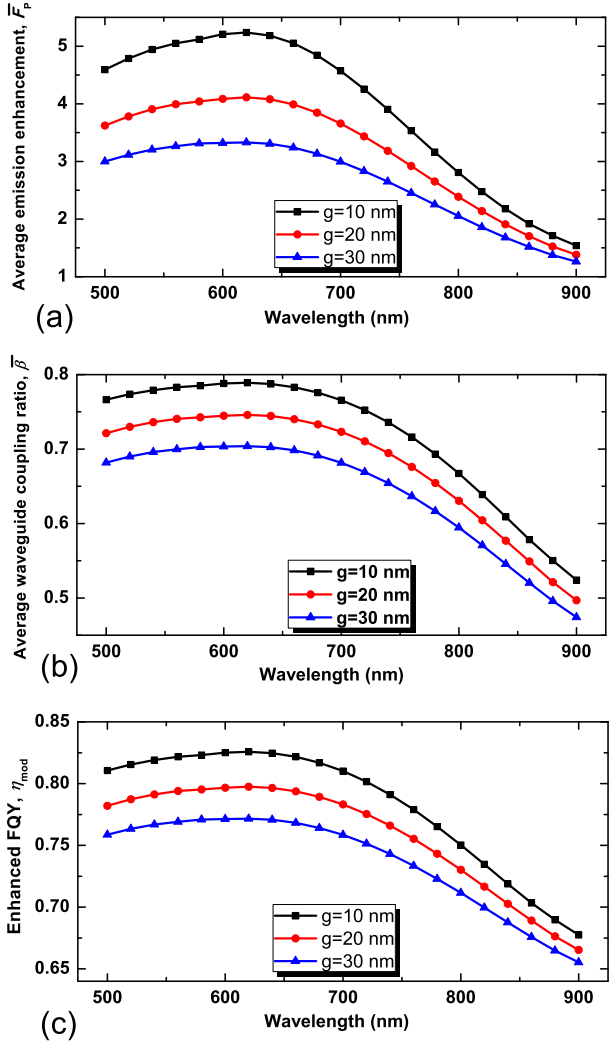}}\caption{(a),
(b) Average spontaneous emission enhancement ($\overline{F}_{p}$) and average
waveguide coupling efficiency ($\overline{\beta}$) depending on the working
wavelength for slot waveguides, respectively. (c) The enhanced FQY $\eta$ as a
function of wavelength with initial FQY $\eta_{0}=0.5$. In all cases, $a=100$
\textrm{nm}, $b=150$ \textrm{nm}.}%
\end{figure}

In the analysis above, we have assumed that the dipole is oriented parallel to
the electric field of the quasi-TE mode. In an actual case, the orientation of
a dipole may be isotropically distributed over any direction. In general, the
quasi-TE mode has a much stronger electric field concentration in the slot
region than the quasi-TM mode. As a result, the dipole oscillating in the $y$
direction (coupled to the quasi-TE mode) exhibits a much larger emission
enhancement than the dipole oscillating in the $x$ direction (coupled to the
quasi-TM mode). In addition, the spontaneous emission enhancement of the
dipole oscillating in the $z$ direction is even smaller because both the modes
are quasi-transverse. The isotropically averaged emission enhancement factor
$\overline{F}_{p}$ can be calculated by%
\begin{equation}
\overline{F}_{p}=\frac{F_{p,x}+F_{p,y}+F_{p,z}}{3}, \label{eq8}%
\end{equation}
where $F_{p,i=x,y,z}$ denotes the spontaneous emission enhancement when the
dipole oscillates in the $i$ direction. With $\overline{F}_{p}$, the
isotropically averaged waveguide coupling efficiency $\overline{\beta}$ is
thus obtained as%
\begin{equation}
\overline{\beta}=\frac{\overline{F}_{p}}{\overline{F}_{p}+n_{1}}. \label{eq9}%
\end{equation}

Figures 4(a) and 4(b) show $\overline{F}_{p}$ and $\overline{\beta}$,
respectively, depending on the emission wavelength for three different slot
widths $g=10,20,30$ nm. From Figure 4(a), on one hand, due to enhanced
dipole-field coupling, $\overline{F}_{p}$ increases as the slot width
decreases. On the other hand, $\overline{F}_{p}$ still keeps high over the
whole calculation range from $500$ to $800$ nm. Thus, even if dipoles
oscillate in random directions, most emission of the dipoles will couple into
the waveguide modes. For instance, as demonstrated in Figure 4(b), the
averaged waveguide coupling efficiency $\overline{\beta}$ is high above $0.7$
over a broad wavelength range in the case of $g=10$ nm.

It should also be noted that only spontaneous emission rate is speeded up by
the concentrated slot mode, while the intrinsic nonradiative decay path
related to the initial FQY $\eta_{0}$ keeps unchanged, leading to an enhanced
FQY \cite{Purcell}. The discussion above has assumed a unit initial FQY
($\eta_{0}=1$). For a finite FQY, however, we find that the spontaneous
emission enhancement can actually enhance the FQY. In the slot waveguide, the
modulated (enhanced) FQY $\eta_{\text{\textrm{mod}}}$ describing the ratio of
the radiation, can be calculated by%
\begin{equation}
\eta_{\text{\textrm{mod}}}=\frac{\eta_{0}(1+\overline{F}_{p}/n_{1})}%
{(1-\eta_{0})+\eta_{0}(1+\overline{F}_{p}/n_{1})}. \label{eq10}%
\end{equation}
As demonstrated in Figure 4(c), for an initial $\eta_{0}=0.5$, the modified
FQY $\eta_{\text{\textrm{mod}}}$ exceeds $0.8$ over a broad wavelength range
in the case of $g=10$ \textrm{nm}.

We turn to analyze the total photon conversion efficiency of this LSC. Besides
the isotropically averaged waveguide coupling efficiency $\overline{\beta}$
and the modified FQY $\eta_{\text{\textrm{mod}}}$, the total efficiency
$\eta_{\text{\textrm{LSC}}}$ should also include the solar photon absorption
efficiency $\eta_{\text{\textrm{abs}}}$ by the active medium, the
transportation efficiency $\eta_{\text{\textrm{tran}}}$\ from emitters to the
PV systems. Thus, we have%
\begin{equation}
\eta_{\mathrm{LSC}}=\eta_{\text{\textrm{abs}}}\eta_{\text{\textrm{mod}}%
}\overline{\beta}\eta_{\mathrm{tran}}. \label{eq11}%
\end{equation}

In the following we will explain how the slot waveguides play significant
roles in obtaining a high $\eta_{\mathrm{LSC}}$. First, $\eta
_{\text{\textrm{abs}}}$ depends on the intrinsic property of the active
medium, such as the absorption cross-section and the absorption spectrum.
Interestingly, the absorption ability of single emitters is expected to be
improved due to the enhanced emission with a high $\overline{F}_{p}$ in the
slot waveguide. A higher $\overline{F}_{p}$ corresponds to a faster
spontaneous emission of the active medium molecules, which implies the
stimulated molecules can go back to their ground state faster and the round
time of the absorption-emission process is significantly shortened
\cite{Purcell}. In other words, the absorption length of the slot region is
efficiently increased. Second, in Figure 4, both $\eta_{\text{\textrm{mod}}}%
$\ and $\overline{\beta}$\ are improved over a broad wavelength range, which
can be further improved by aligning the dipoles of luminescent centers
parallel to the electric field of quasi-TE modes \cite{Rowan,2009AFM}. Third,
the luminescent photons should be transported to the LSCs edges where the PV
systems can absorb these photons and convert them to photoelectrons. In this
process, the photon transportation efficiency $\eta_{\text{\textrm{tran}}}$
can be high because the slot provides good propagating modes along $z$
direction. Nevertheless, $\eta_{\text{\textrm{tran}}}$ may be degraded by the
re-absorption phenomenon of the active medium. Potentially, the re-absorption
can be suppressed by choosing luminescent centers whose absorption and
emission spectra have a small overlap due to a large Stokes shift
\cite{STOKES}, or using luminophores with unitary FQY which effectively close
the nonradiative decay channel.

\begin{figure}[ptbh]
\centerline{\includegraphics[keepaspectratio=true,width=0.45\textwidth]{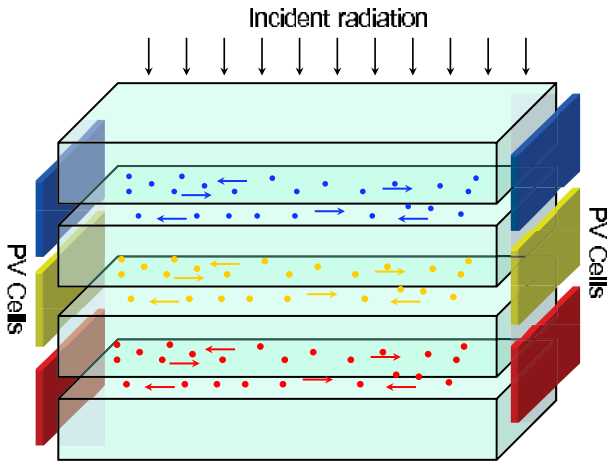}}\caption{Schematic
illustration of a tandem slot waveguide structured LSC. From top to
bottom,
the slot waveguides absorb solar photons with increasing wavelengths}%
\end{figure}

Finally, it should be noticed that slot waveguides possess an optimized
geometrical size for a specific wavelength band. For instance, if the
thickness $a$\ of high-index layers is too large, more energy will distribute
in the high-index layers and less in the slot waveguide, so the Purcell effect
of this structure will be dramatically weakened. As an example, if we increase
the size $a\times b$ to $3a\times3b$, for $550$ nm wavelength in Figure 3(a),
$F_{p}$ decreases from $11$ to $1.4$, and $\beta$ from $0.9$ to $0.5$%
\textit{.} Nevertheless, an expanded geometry $a\times b$ is required to
support propagating modes well in the slot waveguide for long wavelengths. To
obtain the highest power efficiencies, on one hand, the absorption band can be
broadened by mixing dyes and quantum dots with different sizes. On the other
hand, it is also potential to construct tandem LSCs which can absorb more
solar photons in a broad band. Incident solar photons are first absorbed by a
LSC employing a short-wavelength active medium, such as a specific dye.
Photons with longer wavelengths are transmitted through the first LSC and then
absorbed by the second LSC employing a long-wavelength active medium and an
expanded geometry $a\times b$, as shown in Figure 5. To fabricate such a LSC
structure, a nanoimprinting technique can be utilized which can reach large
areas up to several inches wide \cite{ACSnano}.

In summary, we present a theoretical description of nanoscale LSCs where the
conventional ray-optics model is no longer applicable. Based on the
Fermi-golden rule, we evaluate the spontaneous emission enhancement $F_{p}$
and the waveguide coupling efficiency $\beta$ in a slot waveguide consisting
of a nanometer-sized low-index slot region sandwiched by two high-index
regions. It is found that the slot waveguide provides a broadband enhancement
of light harvesting by the luminescence centers in the slot region. In spite
of a low initial FQY $\eta_{0}=0.5$, approximately $80\%$ solar photons can
re-emit as luminescent photons, and more than $80\%$ luminescence can be
coupled into the waveguide modes for ultimate absorption by the solar cell
located on the waveguide edge. This LSCs scheme holds a great potential to

construct a tandem structure which can absorb nearly full-spectrum solar
photons, and is of special interest for building integrated nano-solar-cell applications.

\textbf{Acknowledgment.} The authors acknowledge financial support from the
National Natural Science Foundation of China under Grant No. 10821062 and
11004003, the National Basic Research Program of China under Grant Nos.
2007CB307001 and 2009CB930504. Yun-Feng Xiao was also supported by the
Research Fund for the Doctoral Program of Higher Education (No.20090001120004)
and the Scientific Research Foundation for the Returned Overseas Chinese Scholars.


\begin{thebibliography}{99}                                                                                               %


\bibitem {nano3}W. U. Huynh, J. J. Dittmer, A. P. Alivisatos, "Hybrid
nanorod-polymer solar cells," Science, vol. 295, pp. 2425-2427, 2002.

\bibitem {nano4}M. Law, L. E. Greene, J. C. Johnson, R. Saykally, P. Yang,
"Nanowire dye-sensitized solar cells," Nature Material, vol. 4, pp. 455-459, 2005.

\bibitem {nano8}D. H. Ko, J. R. Tumbleston, L. Zhang, S. Williams, J. M.
DeSimone, R. Lopez, E. T. Samulski, "Photonic Crystal Geometry for Organic
Solar Cells," Nano Lett., vol. 9, pp. 2742-2746, 2009.

\bibitem {nano12}J. G. Mutitu et al., "Thin film solar cell design based on
photonic crystal and diffractive grating structures," Opt. Express, vol. 16,
pp. 15238-15248, 2008.

\bibitem {t2}A. Aubry et al., "Plasmonic light-harvesting devices over the
whole visible spectrum," Nano Lett., vol. 10, pp. 2574-2579, 2010.

\bibitem {t5}L. Y. Cao et al., "Semiconductor Nanowire Optical Antenna Solar
Absorbers," Nano Lett., vol. 10, pp. 439-445, 2010.

\bibitem {t6}W. Wang, S. Wu, K. Reinhardt, Y. Lu, S. Chen, "Broadband light
absorption enhancement in thin-film silicon solar cells," Nano Lett., vol. 10,
pp. 2012-2018, 2010.

\bibitem {jason}J. H. Karp, E. J. Tremblay, J. E. Ford, "Planar micro-optic
solar concentrator," Opt. Express, vol. 18, pp. 1122-1133, 2010.

\bibitem {weber}W. H. Weber, J. Lambe, "Luminescent greenhouse collector for
solar radiation," Appl. Opt., vol. 15, pp. 2299-2300, 1976.

\bibitem {batchelder}J. S. Batchelder, A. H. Zewail, T. Cole, "Luminescent
solar concentrators 2: Experimental and theoretical analysis of their possible
efficiencies," Appl. Opt., vol. 20, pp. 3733-3754, 1981.

\bibitem {AP1977}A. Goetzberger, W. Greubel, "Solar energy conversion with
fluorescent collectors," Appl. Phys., vol. 14, pp. 123-139, 1977.

\bibitem {michael}M. J. Currie, J. K. Mapel, T. D. Heidel, S. Goffri, M. A.
Baldo, "High-Efficiency Organic Solar Concentrators for Photovoltaics,"
Science, vol. 321, pp. 226-228, 2008.

\bibitem {IEEE}B. C. Rowan, L. R. Wilson, B. S. Richards, "Advanced Material
Concepts for Luminescent Solar Concentrators," IEEE J. Selected Topics in
Quantum Electron., vol. 14, pp. 1312-1322, 2008.

\bibitem {Chatten}A. J. Chatten, K. W. J. Barnham, B. F. Buxton, N. J.
Ekins-Daukes, M. A. Malik, "Quantum Dot Solar Concentrators," Semiconductors,
vol. 38, pp. 609-617, 2004.

\bibitem {Earp}A. A. Earp, G. B. Smith, P. D. Swift, J. Franklin, "Maximising
the light output of a Luminescent Solar Concentrator," Sol. Energy, vol. 76,
pp. 655-667, 2004.

\bibitem {Carrascosa}M. Carrascosa, S. Unamuno, F. Agullo-Lopez, "Monte Carlo
simulation of the performance of PMMA luminescent solar collectors," Appl.
Opt., vol. 22, pp. 3236-3241, 1983.

\bibitem {Rau}U. Rau, F. Einsele, G. C. Glaeser, "Efficiency limits of
photovoltaic fluorescent collectors," Appl. Phys. Lett., vol. 87, pp. 171101, 2003.

\bibitem {OL2004}V. R. Almeida, Q. F. Xu, C. A. Barrios, M. Lipson, "Guiding
and confining light in void nanostructure," Opt. Lett., vol. 29, pp.
1209-1211, 2004.

\bibitem {OL2007}C. A. Barrios et al., "Slot-waveguide biochemical sensor,"
Opt. Lett., vol. 32, pp. 3080-3082, 2007.

\bibitem {OE2009}Y. C. Jun. R. M. Briggs, H. A. Atwater, M. L. Brongersma,
"Broadband enhancement of light emission in silicon slot waveguides," Opt.
Express, vol. 17, pp. 7479-7490, 2009.

\bibitem {Purcell}P. Goy, J. M. Raimond, M. Gross, S. Haroche, "Obervation of
cavity-enhanced single-atom spontaneous emission," Phys. Rev. Lett., vol 50,
no. 24, pp. 1903-1906, 1983.

\bibitem {Rowan}R. W. MacQueen, Y. Y. Cheng, R. G. C. R. Clady, T. W. Schmidt,
"Towards an aligned luminophore solar concentrator," Opt. Express, vol. 18,
pp. A161-A166, 2010.

\bibitem {2009AFM}Paul P. C. Verbunt, A. Kaiser, K. Hermans, C. W. M.
Bastiaansen, D. J. Broer, M. G. Debije, "Controlling Light Emission in
Luminescent Solar Concentrators Through Use of Dye Molecules Aligned in a
Planar Manner by Liquid Crystals," Adv. Funct. Mater., vol. 19, pp. 2714-2719, 2009.

\bibitem {STOKES}L. R. Wilson, B. C. Rowan, N. Robertson, O. Moudam, A. C.
Jones, B. S. Richards, "Characterization and reduction of reabsorption losses
in luminescent solar concentrators," Appl. Optics, vol. 49, pp. 1651-1661, 2010.

\bibitem {ACSnano}S. Ahn, L. Guo, "Large-Area Roll-to-Roll and Roll-to-Plate
Nanoimprint Lithography: A Step toward High-Throughput Application of
Continuous Nanoimprinting," ACS Nano, vol. 3, pp. 2304-2310, 2009.
\end{thebibliography}
\end{document}